# NEURAL SOURCE/SINK PHASE CONNECTIVITY IN DEVELOPMENTAL DYSLEXIA BY MEANS OF INTERCHANNEL CAUSALITY

I. RODRÍGUEZ-RODRÍGUEZ, A. ORTIZ, N.J. GALLEGO-MOLINA, M.A. FORMOSO
*Departamento de Ingeniería de Comunicaciones, Universidad de Málaga, 29004 Málaga, Spain*
*{ignacio.rodriguez, aortiz, njgm, marco.a.formoso}@ic.uma.es*

W. L. WOO
*Department of Computer and Information Sciences, Northumbria University, Newcastle upon Tyne NE1 8ST, UK*
*wailok.woo@northumbria.ac.uk*

While the brain connectivity network can inform the understanding and diagnosis of developmental dyslexia, its cause-effect relationships have not yet enough been examined. Employing electroencephalography signals and band-limited white noise stimulus at 4.8 Hz (prosodic-syllabic frequency), we measure the phase Granger causalities among channels to identify differences between dyslexic learners and controls, thereby proposing a method to calculate directional connectivity. As causal relationships run in both directions, we explore three scenarios, namely channels' activity as sources, as sinks, and in total. Our proposed method can be used for both classification and exploratory analysis. In all scenarios, we find confirmation of the established right-lateralized Theta sampling network anomaly, in line with the temporal sampling framework's assumption of oscillatory differences in the Theta and Gamma bands. Further, we show that this anomaly primarily occurs in the causal relationships of channels acting as sinks, where it is significantly more pronounced than when only total activity is observed. In the sink scenario, our classifier obtains 0.84 and 0.88 accuracy and 0.87 and 0.93 AUC for the Theta and Gamma bands, respectively.

## 1. Introduction

Developmental dyslexia (DD) is a learning difficulty that typically causes various reading difficulties, including letter migration and frequent spelling errors. In any given population, between 5% and 12% of learners are likely to have DD, depending on the test battery used [1]. DD is traditionally diagnosed using behavioral tests of reading and writing skills, but these are vulnerable to exogenous factors, such as attitude or disposition, leading to diagnoses that may be fundamentally unsound [2]. It is, therefore, imperative to develop more objective metrics that can offer a more accurate diagnosis among young learners. A stimulus system that remains uninfluenced by the learner's behavior, actions and context (e.g., native language or learning level) would be extremely valuable. If the stimulus further involves the simulation of prosody, i.e. the white noise at the usual frequency of the language envelope, it may also inform our understanding of the brain areas active in auditory processing, indicating the differences between learners with and without dyslexia.

While various neuroscience methods for gathering functional brain data exist, including functional magnetic resonance imaging (fMRI) [3], magnetoencephalography (MEG) and functional near-infrared spectroscopy (fNIRS), electroencephalography (EEG) continues to be the most widely used and least costly method to assess cortical brain activity with enhanced temporal resolution. An EEG measures several frequency bands, namely the Delta, Theta, Alpha, Beta, and Gamma bands, which do not experience stimulation equally, and it is thus generally held that the stimulation of one band can transfer to the others. Using EEG to separately investigate the patterns emerging in these bands may offer valuable insights for the research on DD.

EEG is well-established in DD studies exploring the functional network connectivity and organization of the brain. Functional connectivity means the level of coordination between the activities in different areas of the brain while the learner is engaging in a task. Prior research has produced various techniques that employ EEG to assess functional connectivity to, for example, determine what patterns are characteristic of neurological conditions, including Parkinson's disease [4]. Studies in cognitive neuroscience have also used brain connectivity to identify brain areas crucial to language and learning [5].

Connectivity analysis has allowed neuroscience to provide even deeper insights [6] by analyzing the parameters linking two signals gathered through two distinct channels, such as their correlation, causality and covariance [7]. Employing connectivity analysis to brain signals measured in different regions allows us to explore the neural network, in line with the notion that the brain is hyper-connected [8].

Notably, brain connectivity is not solely limited to the interactions between areas, with regions potentially influencing each other through, e.g., phase-phase or phase-amplitude modulations among bands [9]. We here consider areas that primarily exert an influence as *sources*, while those more likely to be subject to this influence are *sinks*. This novel consideration of connectivity in terms of sources and sinks can not only serve classification but also facilitate exploratory analysis.

We propose extracting the frequency components of each band from the EEG signals acquired under prosodic auditory stimuli, subsequently using them to generate a connectivity model based on inter-channel Granger causality [10]. By modelling connectivity as sources and sinks, we seek to clarify the existence of abnormalities between learners with and without DD, aiming to offer an enhanced understanding of the mechanisms underlying DD, ultimately allowing early diagnosis.

The remainder of the paper is structured as follows. Section 2 describes the most relevant extant work in this field, and Section 3 outlines the data and methodology, including the preprocessing, Granger causality matrices and connectivity matrices construction, and classification algorithms. Section 4 presents our main results, leading to the discussion in Section 5. Finally, Section 6 presents the main conclusions and contributions of this work.

## 2. Related works

Previous studies have indicated that the phonological deficit that causes DD may be due to an impairment in the neural encoding of low-frequency speech envelopes, relating to speech prosody [11]. There is evidence of significant difficulties among that learners with DD in tasks relying on prosodic awareness, e.g. identifying syllable stress, compared to controls at an earlier reading level [12]. This indicates the presence of atypical oscillatory functioning in low-frequency brain rhythms in DD [13].

There has been substantial research on the important role played by the ability to perceive prosodic frequency. After directly measuring the neural encoding of children's speech using EEG, Power et al. [11] reconstructed the participants' speech stimulus envelopes using the emergent patterns. The EEG recordings were done while the participants were performing a word report task using noise-vocoded speech, i.e. still with a low-frequency envelope yet with a degraded temporal fine structure (TFS) of speech. Due to this degradation, the participants necessarily derived the spoken words and sentences from the information given by the envelope. If the learners could accurately perceive the words and sentences, it was possible to evaluate the functioning of their neural encoding of the low-frequency envelopes in speech, which is likely impaired in learners with DD according to temporal sampling theory.

Brain activity, and thus the connectivity network, occurs across various frequency bands, as demonstrated via the temporal sampling framework (TSF). Temporal coding is thought to be partially attributed to synchronous auditory cortex activity, wherein the network neurons synchronize endogenous oscillations at different preferred rates while matching the temporal information of the acoustic speech signal [14] [15] [16]. The auditory and visual parts of speech unfold across different timescales, and thus, when the neurons in auditory and visual cortices oscillate, they are believed to phase-align their activity to match the input's modulation rates [17].

TSF proposes that atypical oscillatory sampling at various temporal rates may be the cause of the phonological impairment in DD. Furthermore, a potential biological mechanism for DD has recently been suggested, highlighting the presence of atypical dominant neural entrainment [18] for the slow rhythmic prosodic (0.5–1 Hz), syllabic (4–8 Hz) and phoneme (12–40Hz) rhythm categories [19]. Following this line of thought, we might consider learners with DD to have atypical oscillatory sampling for at least one temporal rate, leading to difficulties in phonologically capturing linguistic units such as syllables or phonemes.

However, this phenomenon is not likely to be experienced equally across all frequency bands (i.e. Delta, Theta, Alpha, Beta, and Gamma). Thus, it seems pertinent to examine these bands' connectivity patterns separately using EEG. Prior research has indeed used EEG or MEG to investigate the fundamental mechanisms underlying DD, implementing speech-based stimuli under the premise that DD is essentially derived from a lesser awareness of individual speech units [20]. Using visual and auditory stimulus, Power et al. [21], for example, identified differences between learners with DD and a control group in the preferred entrainment phase of the Delta and Theta bands. Based on changes in the frequency, phase, and power spectrum, it thus becomes feasible to derive measures of spectral connectivity. In line with this, there are techniques showing the statistical relationship between electrodes on the same frequency band [22].

The prior research has also explored the inference from connectivity patterns during reading tasks. For example, Žarić et al. [23] used visual word and false font processing tasks to investigate disruptions in the connectivity between the visual and language processing networks. They hereby calculated the connectivity patterns based on how statistically significant the differences in the power spectral density (PSD) were for each EEG band. Language-based reading or writing-related tasks have also been used in previous studies identifying discriminant patterns in EEG signals. For instance, using graph theory, González et al. [24] compared the EEG measurements of participants performing audiovisual tasks or at rest to determine differences in the connectivity patterns. Meanwhile, Stam et al. [25] used a phase lag index to compute multiple weighted connectivity matrices for multiple frequency bands.

Assessing the connectivity of two channels requires a separate analysis of their respective phases. A signal's phase, $\varphi(t)$, changes over time when being captured with an electrode, and thus it must be measured for each channel $i$, referring to the instantaneous phase, captured using a Hilbert transform and computed via band-pass filtered signals. Consequently, the phase value can be pinpointed at each time point, allowing inter-channel correlation and causality to be determined. Using this method to track changes in the phase synchronization of epileptic patients, Mormann et al. [26] showed that epileptic episodes are often preceded by characteristic changes in synchronization.

Following this, we can estimate the inter-channel connectivity based on the cause-effect relationships. The Granger causality test can hereby show whether one of the factors is a time series, allowing the characteristics of additional time series to be predicted. First employed in the 1980s in the economics field, Granger causality is a statistical hypothesis test that has been used to produce good results in a wide range of other fields [10]. Neuroscience research has applied it to EEG measurements, producing findings on brain activity in emotion recognition [27], Vagus nerve stimulation [28], and pain perception [29].

Connectivity based on causality implies cause-effect relationships between various areas of the brain, but these are not necessarily bidirectional. Thus, some brain areas will be very active because they are influencing others, and other areas may be very active because they are being influenced by remote areas. Likewise, it could be the case that high activity may be due to both situations. While this concept of sources/sinks is not new, it has been subject to a variety of different approaches. For example, Rimehaug et al. [30] integrated it into their model of the visual cortex's local field potential, while Sotero et al. [31] used it to explain the laminar distribution of phase-amplitude coupling of spontaneous current sources and sinks in rat brains. However, neither of those studies based their modeling of sources and sinks on causality relationships, instead using the electrical activity in the cerebral cortex.

The concepts of Granger causality and source/sink relationships have been used to address the clinical issue of surgical resection planning by capturing high-frequency ictal and preictal oscillations on an intracranial EEG [32], although no connectivity maps were constructed; furthermore, the study did not use machine learning to examine whether this approach could be applied in the differential diagnosis of impairments.

Building on the work outlined above, we apply machine learning classification algorithms to assess the potential of diagnosing DD via a learner's sources, sinks and total activity under stimulus, identified using Granger causality matrices. Due to the impenetrable nature of EEG signal classification and the complexity of the problem being addressed, machine learning is highly suitable [33]. Briefly, we seek to demonstrate that different connectivity patterns are induced in certain brain networks by low-level auditory processing. To this end, we delineate this connectivity by establishing the source and sink relationships through the application of Granger causality to the phase synchronization among EEG channels.

## 3. Materials and methods

### 3.1. *Data acquisition*

The dataset comprised EEG data from the University of Málaga's Leeduca Study Group [34], gathered from 48 age-matched child participants (32 skilled readers and 16 dyslexic readers) ($t(1) = -1.4$, $p > 0.05$, age range: 88-100 months). All participants were righthanded native Spanish speakers with normal or corrected-to-normal vision; none had a hearing impairment. All participants in the dyslexic group had been formally diagnosed with dyslexia at school. All participants in the skilled reader group were free from reading and writing difficulties and had not been formally diagnosed with dyslexia. The participants' legal guardians expressed their understanding of the study, gave their written consent, and were present throughout the experiment.

All participants experienced an auditory stimulus in 15-minute sessions. The stimulus, which was modulated at 4.8 Hz (prosodic-syllabic frequency) in 2.5-minute segments, was band-limited white noise. This type of stimulus was chosen to identify what synchronicity patterns the low-level auditory processing would induce and on the basis of the expert knowledge of linguistic psychologists concerning the main frequency components representing words in the human voice. The participants' EEG signals were recorded with a BrainVision actiCHamp Plus with 32 active electrodes (actiCAP, Brain Products GmbH, Germany) at a 500 Hz sampling rate. The 10–20 standardized system was used to place the 32 electrodes.

### 3.2. *Preprocessing*

The preprocessing involved removing all eye-blinking and movement/impedance variation artifacts from the EEG signals. The former were eliminated via independent component analysis (ICA) [35] based on the eye movements observed in the EOG channel, while for the latter the relevant EEG segments were excluded. The channels were then referenced to the Cz channel.

Then, a band-pass filter was applied to the EEG channels to collect information for the five EEG frequency bands (Delta, 1.5–4 Hz; Theta, 4–8 Hz, Alpha, 8–13 Hz; Beta, 13–30 Hz; and Gamma, 30–80 Hz). We used finite impulse response (FIR) filters because these ensure a constant phase lag that can later be corrected. To be specific, each signal was sent forward and backward through the two-way zero-phase lag band-pass FIR least-squares filter, producing a zero-lag phase in the overall filtering process that addressed the issue of phase lag [36]. As low-pass filtering with an 80 Hz threshold was employed, we added a 50 Hz notch filter during preprocessing to eliminate this frequency component.

### 3.3. *Hilbert Transform*

A Hilbert transform (HT) transforms real signals into analytic signals, i.e. complex-valued time series without negative frequency components, allowing the time-varying amplitude, phase and frequency, i.e., the instantaneous amplitude, phase and frequency, to be calculated from the analytic signal.

We define HT for a signal $x(t)$ as:

$$\mathcal{H}[x(t)] = \frac{1}{\pi}\int_{-\infty}^{+\infty} \frac{x(t)}{t-\tau} d\tau \quad (1)$$

and we obtain the analytic signal $z_i(t)$ for signal $x(t)$ as:

$$z_i(t) = x_i(t) + j\mathcal{H}\{x_i(t)\} = a(t)e^{j\phi(t)} \quad (2)$$

From $z_i(t)$, computing the instantaneous amplitude is straightforward:

$$a(t) = \sqrt{re(z_i(t))^2 + im(z_i(t))^2} \quad (3)$$

with the instantaneous, unwrapped phase as:

$$\phi(t) = tan^{-1}\frac{im(z_i(t))}{re(z_i(t))} \quad (4)$$

The above technique gives the phase value for each time point, allowing the inter-channel synchronization to be estimated based on the phase variation.

### 3.4. *Granger Causality test*

Developed for the field of econometrics by Clive Granger, Granger causality [37] describes causal interactions occurring between continuous-valued time series. As a statistical hypothesis test, it essentially states that "the past and present may cause the future, but the future cannot cause the past"; hence, knowing a cause will be more helpful in predicting future effects than an auto-regression will. Specifically, variable $x$ will Granger-cause $y$ if the auto-regression for $y$ that uses past values of $x$ and $y$ is significantly more accurate than one using only past values of $y$. We may exemplify this by taking two stationary time-series sequences, $x_t$ and $y_t$, whereby $x_{t-k}$ and $y_{t-k}$ are, respectively, the past $k$ values of $x_t$ and $y_t$. We then use two regressions to perform Granger causality:

$$\widehat{y}_{t_1} = \sum_{k=1}^{l} a_k y_{t-k} + \varepsilon_t \quad (5)$$

$$\widehat{y}_{t_2} = \sum_{k=1}^{l} a_k y_{t-k} + \sum_{k=1}^{w} b_k x_{t-k} + \eta_t \quad (6)$$

where $\widehat{y}_{t_1}$ and $\widehat{y}_{t_2}$ are, respectively, the fitting values of the first and second regressions; $l$ and $w$ are the maximum numbers of the lagged observations of $x_t$ and $y_t$; $a_k$; $b_k \in$ R are the regression coefficient vectors estimated using least squares; and $\varepsilon_t$ and $\eta_t$ are white noise (prediction errors). Note that even though $w$ can be infinite, due to the finite nature of our data, we consider $w$ finite and give it a length well below the time series length, estimated using model selection, such as the Akaike information criterion (AIC) [38]. Next, an F-test is applied to give a p-value indicating whether the regression model produced by Eq. (5) is statistically better than that of Eq. (6). If it is, then *x* Granger-causes *y*.

We perform Granger causality testing for each participant and evaluate the channels' interactions, producing an *n* x *n* square matrix of p-values (*n* = number of channels).

Using Granger causality to analyze the neural network's directed functional connectivity intuitively demonstrates the directionality with which information is transmitted between neurons or brain regions. Previous studies have already applied this technique to EEG analysis with great success [39][40].

### 3.5. *Connectivity vectors*

The field of neuroscience tends to consider the brain as a network using functional information [41][42][43], culminating in the so-called connectome. This refers to the complete mapping of all connections between brain regions as an adjacency matrix, and often includes the covariance, as well as other metrics, between fMRI signals measured for different regions. Several studies have also examined the temporal covariance between EEG electrodes.

Once we had assembled the Granger causality matrices for each participant subject, we established a threshold value that evidenced a causal relationship between the channels. Then, we formulated the three scenarios used to produce each participant's feature set:
- *Sources*: Array of *n* x 1 elements; each element relates each channel with the number of channels that it influences.
- *Sinks*: Array of *n* x 1 elements; each element relates each channel with the number of channels that it is influenced by.
- *Total activity*: Array of *n* x 1 elements; the sum of the two previous scenarios, acting as a reference for each channel's global activity.

By organizing the information thus, we receive the same number of features as there are channels for each participant, each with a number that indicates its activity as a source, as a sink, or the total. A summary of this process is presented in Figure 1.

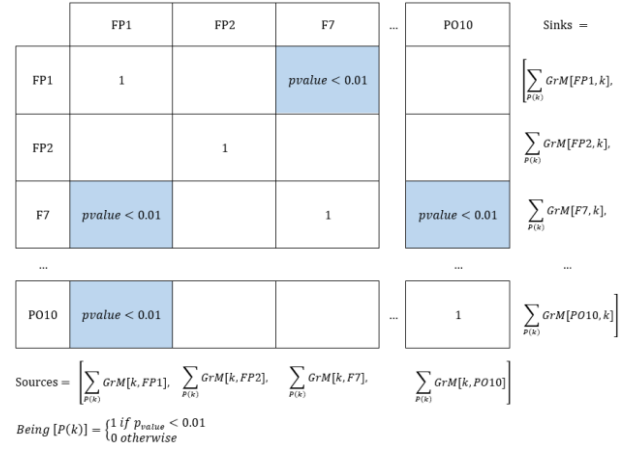

Fig. 1. Assembling the source and sink connectivity arrays for a participant, given the relevant Granger matrix. [P(k)] is an Iverson bracket function.

### 3.6. *Ensemble feature selection*

If the model includes many features, it will be more complex, potentially leading to data overfitting. Moreover, some of the features may be noise and could adversely affect the model. Thus, we removed such features to ensure the better generalization of the model. We hereby selected the variables based on majority voting through the application of several techniques. If a variable was chosen by an algorithm, it received one vote. The votes were then summed for each variable, and those with the most votes were selected. (Fig. 2). This method has been found to be suitable for datasets that are high-dimensional yet have few instances [44]. The voting strategy used a variety of feature selection methods [45], as outlined in the following:

**Information value (IV) using weight of evidence (WOE):** This indicates the predictive power of an independent variable concerning the dependent variable [46]. It allows a continuous independent variable to be transformed into a set of groups or bins based on the similarity of the dependent variable distribution (i.e. numbers of events and non-events). Using WOE allows outliers and missing values to be addressed and eliminates the need for dummy variables [47]:

$$WOE = \ln\left(\frac{Event\%}{Non\ Event\%}\right) \quad (7)$$

$$IV = \Sigma[(Event\% - Non\ Event\%) * WOE] \quad (8)$$

An IV statistic above 0.3 is held to indicate a strong relationship between the predictor and the event/non-event odds ratio [48].

**Variable importance using random forest/extra trees classifier:** Calculated using a tree-based estimator, this can be used to eliminate irrelevant features. Variable importance is conventionally computed using the *mean decrease in impurity* (i.e., *gini importance* [49]) mechanism, wherein the improvement in the split criterion for each split of each tree is the importance measure assigned to the splitting variable. For each variable, this is separately accumulated over all the trees in the forest. This measure is similar to the $R^2$ in the training set regression.

**Recursive Feature Elimination:** This can be used to select features by recursively considering feature sets with diminishing size based on an external estimator (a linear regression model) that assigns weights to the features [50]. The estimator is trained on the first feature set, noting each feature's importance based on a given attribute. The least important features are subsequently removed from the current set. The process is performed recursively on the pruned set until the desired number of features is achieved.

**Chi-square best variables:** This uses a chi-square ($\chi^2$) test to assess the correlations among a dataset's features and identify multicollinearity. The aim is revealing any relationships between the dependent variable and any of the independent variables [51]. In the chi-square test, $H_0$ (null hypothesis) assumes that two features are independent, while $H_1$ (alternative hypothesis) predicts that they are related. We set a $\alpha=0.05$ and a p-value of 0.05 or greater is considered critical, anything less means the deviations are significant hence the hypothesis must be rejected.

**L1-based feature selection:** Some features can be eliminated using a linear model with an L1 penalty. This method involves regularization, wherein a penalty is added to various parameters of a machine learning model to reduce the model's freedom and prevent overfitting. When regularizing linear models, the penalty is applied in addition to the coefficients multiplying the predictors [52]. Unlike other forms of regularization, L1 can reduce some coefficients to zero, meaning the feature is removed.

Once the best variables had been chosen by voting, we performed a multicollinearity check on them.

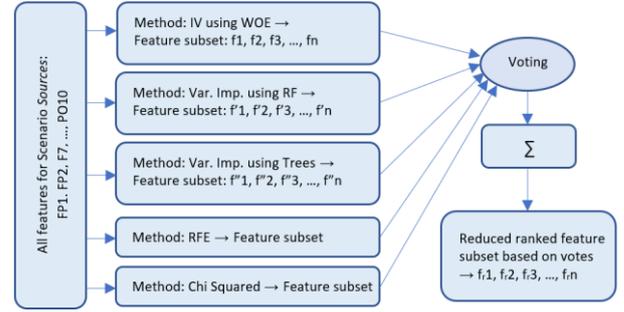

Fig. 2. The feature selection procedure for the 'Sources' scenario using a vote-based approach.

### 3.7. *Classification process*

In an ensemble method, multiple models are first generated and then integrated to produce higher-quality results. The respective predictions are hereby combined using weighted majority voting to make the final prediction. At each boosting iteration, the data are modified by applying $w_1, w_2, ..., w_n$ to each training sample. As the weights are initially $w_i=1/N$, a weak learner is trained in the first step using the raw data. At each successive iteration, the sample weights are modified individually, and the algorithm is then applied to the reweighted data. Training examples that are incorrectly predicted relative to the previous step's boosted model are given increased weights; correctly predicted examples are given decreased weights. As a result, the examples that were difficult to predict become increasingly influential as the number of iterations increases, and the weak learners that follow are forced to focus on the examples previously missed.

Ensemble methods deliver more accurate results than single models, and are particularly suitable for improving binary prediction on small data sets. We use the Gradient Boosting classifier, as well as an Ada Boost for results verification. This latter classifier [53] is a meta-estimator that initially fits to the data, with further copies then being fit to the same data, while incorrectly classified instances' weights are modified to force subsequent classifiers to focus on them. The Gradient Boosting classifier [54] creates an additive model based on a forward stage-wise construction, allowing the optimization of the arbitrary differentiable loss function. At each stage, *n* regression trees are fit to the multinomial or deviance binomial loss function's negative gradient, with a single regression tree being used for the special case of binary classification. To identify the best parameter set, we cross-validate with 20 folds and a parameter grid, as shown in Table 1.

Table 1. Parameter grid of machine learning classifiers.

| Algorithm | Parameter | Range |
|---|---|---|
| Gradient Boosting | n_estimators | 1 to 12 |
| | Loss | deviance, exponential |
| | Learning rate | 0.05 to 1.5 |
| | Criterion | friedm_mse, sq_error, mse, mae |
| | Min_samples_split | 0.01 to 3 |
| | Min_samples_leaf | 0.01 to 3 |
| | Max_depth | 1 to 4 |
| Ada Boost | n_estimators | 1 to 25 |
| | Learning rate | 1 to 3.5 |
| | Boosting algorithm | SAMME, SAMME.R |

## 4. Results

Plotting each learner's array of sources and sinks permits the visual extraction of the respective patterns of the dyslexic and control groups. To this end, we examined the channel distributions for both groups by calculating the means and dispersions and producing a box-and-whisker plot. We also constructed a topoplot as this can illustrate the results with greater clarity. For example, Fig. 3 shows the Theta band connectivity of the control and dyslexic groups specifically for total activity. Please note that Fig. 3 and Fig. 4 do not directly represent the electrical activity of the cerebral cortex, but rather show the levels of the cause-effect relationships between the channels, i.e. in one direction or in the other direction or in total. It immediately becomes clear that despite the similarity of the patterns, the dyslexic group has a significantly higher activity level in the Theta band.

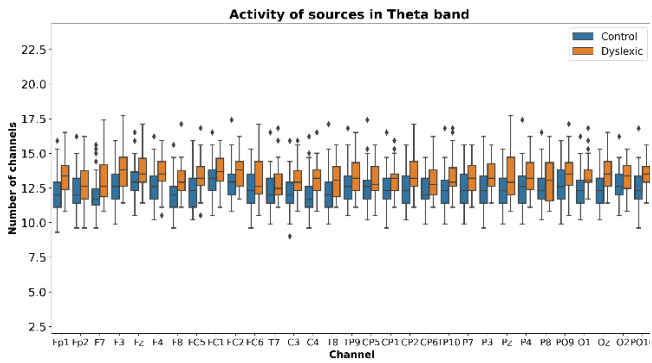

Fig. 3. Boxplot of the total activity in the Alpha band.

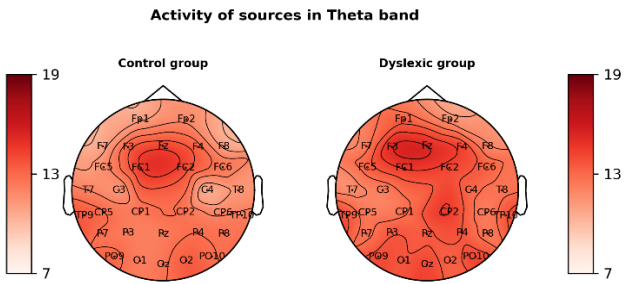

Fig 4. The equivalent graphical representation of Fig. 3 in a topoplot.

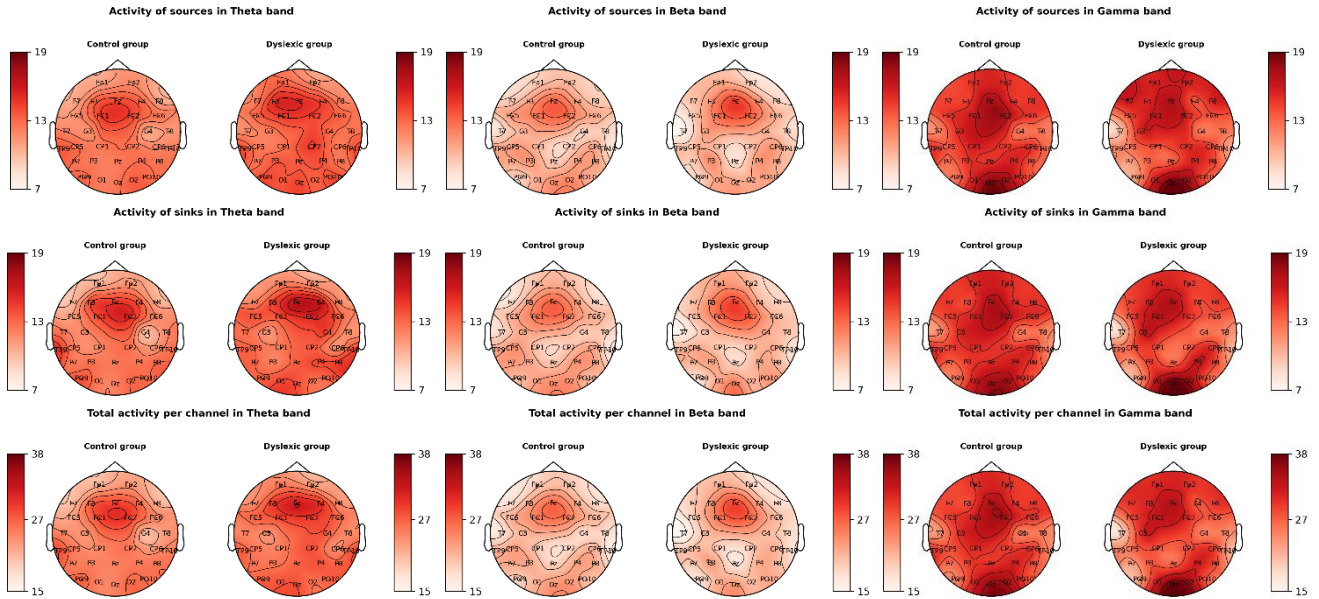

Fig. 5. Source/sink activity in the Theta, Beta and Gamma bands in the control and dyslexic groups. Numbers represent how many channels are affected by each channel as a source, or how many channels are affecting each channel as a sink.

Fig. 5 compares the channel activity in the Theta, Beta and Gamma bands, and can be viewed separately as sources, sinks, or total activity for both the control and dyslexic groups. Please note that the range of visualization is the same in all sinks/sources topoplots, while different in the total activity ones, for better representation. Once more, it is immediately clear that while the patterns are broadly similar, the activity level is higher in the dyslexic group, primarily observed in the sink activity (less in the source activity). Thus, although the sources, broadly speaking, behave similarly between the groups, the dyslexic group has significantly more concentrated sinks and more activity. Consequently, the overall activity level is also affected.

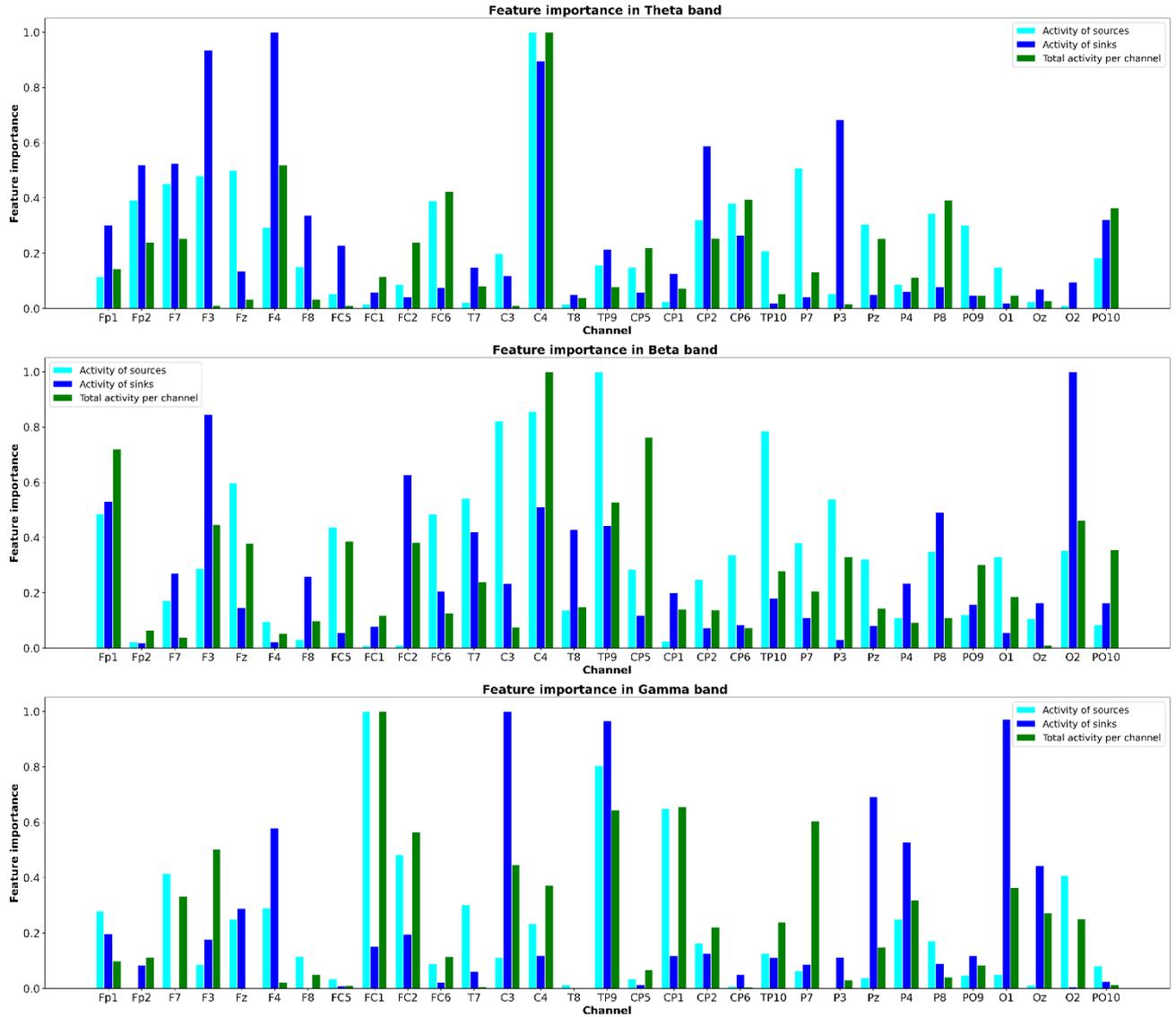

Fig. 6. Feature importance in Theta, Beta and Gamma bands considering sources, sinks and total activity.

With as many arrays as subjects, and with each array having as many components as channels, we performed feature selection to identify channels that can help differentiate between the control and dyslexic groups. The feature selection procedure outlined above was thus applied for the cases of sources, sinks and total activity, according to the band. Fig. 6 presents the results for the Theta, Beta and Gamma bands, whereby the importance values are normalized to permit fair and simple comparison. Channels showing a higher significance are those with more dissimilarity between the control and dyslexic groups, directing us to where we can find different patterns of functioning.

After performing the feature selection for each band, for each case (sources, sinks and total activity), we optimize the Gradient Boosting classifier to obtain the best performance. The results are summarized in Table 2, with performances achieving at least 80% marked bold.

According to the results, the greatest differences between the control and dyslexic groups (i.e., the best classifier results) emerge in the Theta and Gamma bands when accounting for the activity sink role of the different channels, achieving accuracies of 84% and 88%, respectively. We also wish to highlight the results for the Beta band for the activity sources regarding the Area Under the Curve (AUC), in addition to accuracy.

Table 2. Results of the Gradient Boosting machine learning classifier.

| Band | Features set | Accuracy | AUC |
|---|---|---|---|
| Delta | Sources | 0.77 ± 0.14 | 0.65 ± 0.31 |
| | Sinks | 0.79 ± 0.20 | 0.70 ± 0.29 |
| | Total activity | 0.74 ± 0.19 | 0.76 ± 0.25 |
| Theta | Sources | 0.77 ± 0.17 | 0.77 ± 0.30 |
| | **Sinks** | **0.84 ± 0.15** | **0.87 ± 0.18** |
| | Total activity | 0.74 ± 0.17 | 0.72 ± 0.28 |
| Alpha | Sources | 0.79 ± 0.19 | 0.74 ± 0.25 |
| | Sinks | 0.76 ± 0.21 | 0.71 ± 0.29 |
| | Total activity | 0.79 ± 0.17 | 0.77 ± 0.21 |
| Beta | **Sources** | **0.80 ± 0.17** | **0.86 ± 0.18** |
| | Sinks | 0.79 ± 0.24 | 0.81 ± 0.27 |
| | Total activity | 0.76 ± 0.23 | 0.75 ± 0.32 |
| Gamma | Sources | 0.81 ± 0.18 | 0.83 ± 0.22 |
| | **Sinks** | **0.88 ± 0.14** | **0.93 ± 0.16** |
| | Total activity | 0.82 ± 0.12 | 0.87 ± 0.18 |

The Receiver Operating Curve (ROC) space is a valuable data interpretation tool that can be used to assess the performance of a binary classifier, wherein it indicates the cutoff point at which sensitivity is traded for specificity. Hence, it can be used to evaluate the classifier's performance in distinguishing positive and negative samples. Related to this, AUC is the probability that the classifier will assign a random positive instance a more extreme value than a random negative instance. Fig. 7 presents the ROC curves for the Theta, Beta and Gamma bands, to identify those with the best performance. Notably, the Gamma band with the channels' sinks activity as the features presents a 93% under the curve.

The obtained results were verified by repeating the classification process using the Ada Boost algorithm. Table 3 presents the results for the Gamma band while Fig. 8 shows the ROC curve. While the performance is slightly diminished, it remains consistent across all bands and cases (sources, sinks and total activity) with the results from the Gradient Boosting.

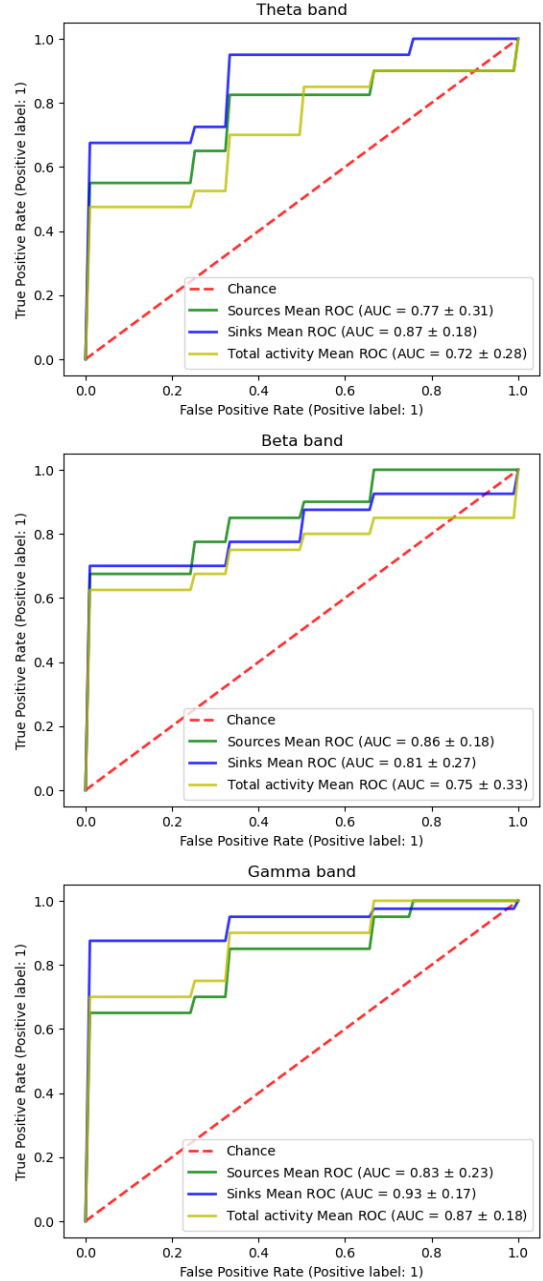

Fig. 7. ROC curves for the Theta, Beta and Gamma bands with the Gradient Boosting classifier.

Table 3. Results for the Ada Boost classifier for the Gamma band.

| Band | Feature set | Accuracy | AUC |
|---|---|---|---|
| Gamma | Sources | 0.83 ± 0.17 | 0.82 ± 0.27 |
| | **Sinks** | **0.88 ± 0.11** | **0.86 ± 0.21** |
| | Total activity | 0.77 ± 0.19 | 0.76 ± 0.31 |

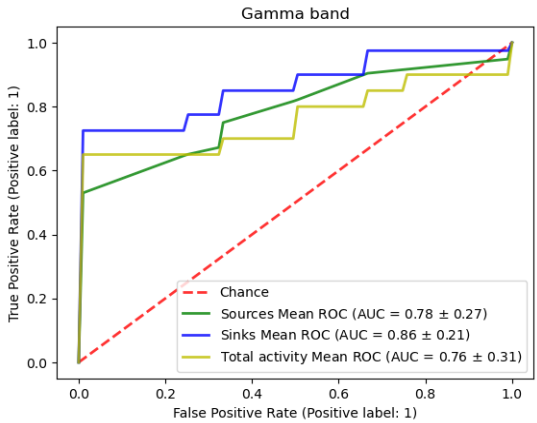

Fig. 8. ROC curves for the Gamma band with the Ada Boost classifier.

As is often the case in biomedical studies, statistical tests are required to check that the number of samples has not introduced bias in the classification stage (e.g., through overfitting). Moreover, there is a need to check the probability of these results having been obtained by chance. For large datasets, such tests need not be as stringent, but real-world studies demand special attention due to the small sample sizes and unbalanced classes. Specifically, in experimental studies the prevalence of the disorder among the population being treated must be taken into account. For DD, this is around 5-12%, as mentioned above.

To this end, a null distribution is generated by estimating the classifier's accuracy for 1000 permutations of the labels. This indicates the distribution for the null hypothesis that the features are not dependent on the labels, and enables the estimation of the probability that the classification results will be reproduced with shuffled labels. The result is an empirical p-value determined by:

$$p-value = \frac{\#perm\ with\ acc.\ higher\ than\ baseline}{\#Number\ of\ permutations} \quad (9)$$

Fig. 9 gives the permutation test results for the Theta, Beta, and Gamma bands for sources, sinks and total activity. The null distribution from the label permutations, as outlined above, is in blue, while the vertical red line represents the accuracy obtained for the non-permuted case. At each permutation iteration, a 20-fold stratified cross-validation is performed, and based on the average of the results obtained at these 20 folds, the corresponding permutation iteration is determined. Hence, Fig. 9 presents the classification's probability density. According to the permutation tests, the results have low p-values and are significant.

## 5. Discussion

The participants were subjected to white noise at 4.8 Hz, i.e. between the syllabic and prosodic frequencies, as the sole stimulus. DD has been shown to link to impairments in syllabic and prosodic perception [55], suggesting general difficulties in identifying the different modulation frequencies. This influences the slower temporal rates of speech processing in particular, as well as the tracking of the amplitude envelope of speech, diminishing learners' syllabic segmentation efficiency.

Multi-time resolution models of speech processing [16] have evidenced that phonetic segment identification associates with faster temporal modulations (Gamma rate, 30–80 Hz), syllable identification is linked to slower modulations (Theta rate, 4–10 Hz), and syllable stress and prosodic patterning information correlates with very slow modulations (Delta rate, 1.5–4 Hz). Nonetheless, anomalies can emerge in various frequency ranges due to inter-band entrainment.

As it offers adequate time resolution, examining the patterns occurring in EEG channels at different bands can unveil the speech encoding linked to problems with speech prosody and sensorimotor synchronization. Exemplifying this, previous research [18] used speech-based stimuli and time-frequency descriptors to reveal the link between speech features and neural dynamics.

We find that the classifier performs better in the Theta and Gamma bands. The results for the Theta band are expected as the TSF suggests that the phonological deficit of DD – regardless of language – may be partially attributed to functionally atypical or impaired phonology entrainment mechanisms in the auditory cortex, especially as oscillations at slower temporal rates, i.e. Theta and Delta, relate to syllabic and prosodic processing [56].

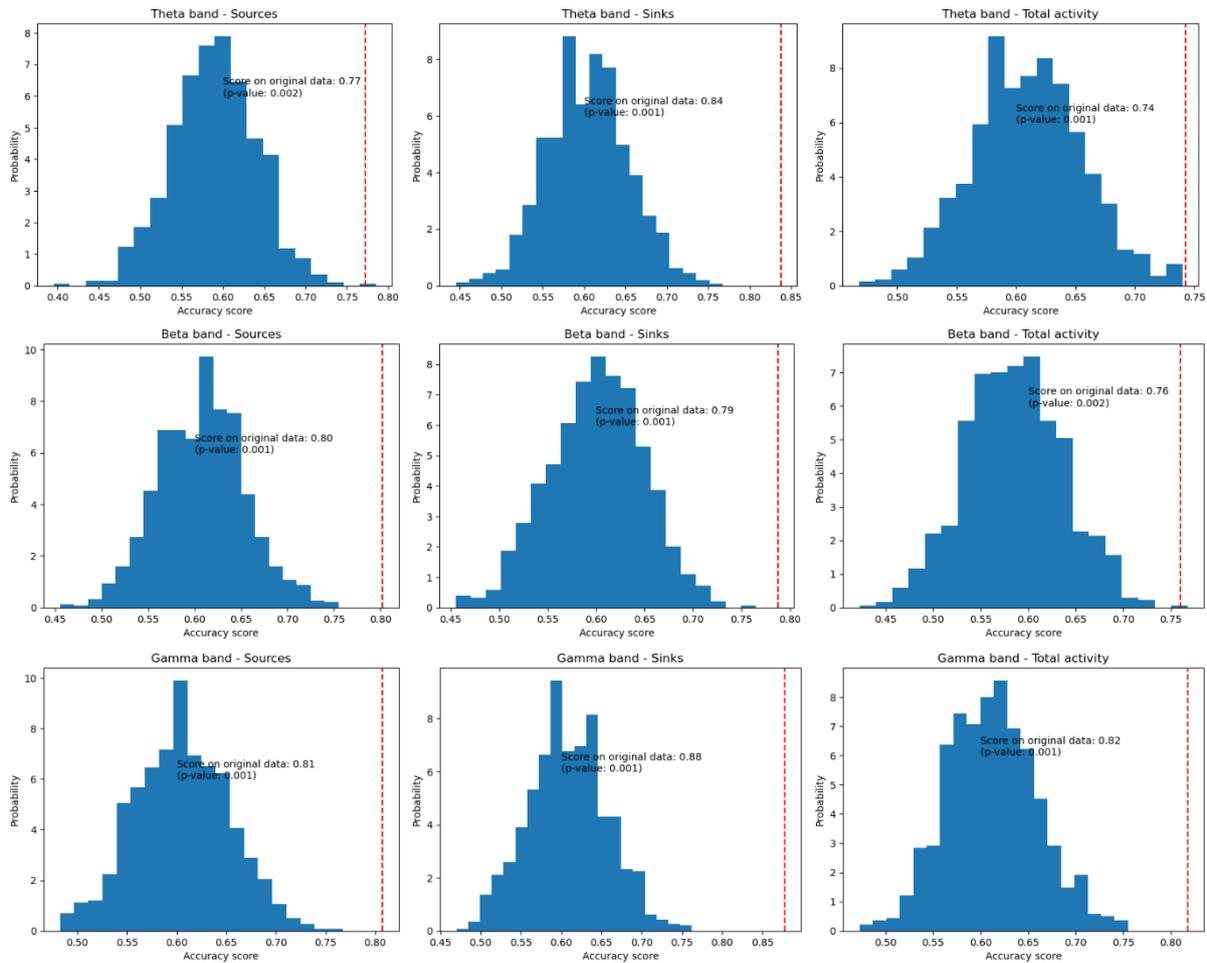

Fig. 9. Permutation tests for Gradient Boosting classifier in Theta, Beta and Gamma bands.

As per the TSF, group differences are expected in neuronal oscillatory entrainment at slower rates (approx. 4 Hz, in line with the stimulus used) [57]. Higher causality relationships emerged in the frontal area in all scenarios for the Theta band. In addition, the number of channels that g-causes causality is higher in the dyslexic domain, which was the case for the sources, sinks and total activity. This higher activity in terms of overall causality relations was evident across all bands. However, in the participants with DD there was significantly less entrainment in the auditory networks of the right hemisphere in the Theta band. As Fig. 6 (feature selection) shows, the C4 channel in the upper part, i.e. the Theta band, is predominantly influential for the causality regarding the sources, as well as the sinks and total activity. It has already been established that the right-lateralized Theta sampling network tends to involve slower temporal rates and codes the speech signal's lower modulation frequencies [57], facilitating syllable-scale temporal integration. In other words, spoken sentences are tracked and distinguished by the Theta band phase pattern, allowing the incoming speech signal to be broken into syllable-sized packets and speech dynamics to be tracked through resetting and sliding, such as with varying rates of speech [58]. Fig. 5 (topoplots) clearly demonstrates that the C4 channel is the most interesting as it has the most Granger causality (causing and being caused) for all scenarios for the dyslexic group. For the sources, the frontal area contains other noteworthy channels (FP2, F7, F3 and Fz) that show differences between the control and dyslexic groups in terms of activity. The most influential channels in the sinks are F3 and F4 (frontal area) and P3.

Hence, it seems pertinent to suggest that the main differences in the causality relationships of the Theta band lie in the so-called dorsal and ventral pathways. In particular, the right area seems critical, as evidenced in the prior research and especially demonstrated here with the sinks scenario.

Another interesting result worth discussing is that for the Beta band. Here, more activity was observed for all three scenarios in the dyslexic group; this agrees with the results for the Theta band as well as those from previous studies [59]. For the sources, differences in the causal relationships were mainly identified in the C3 and C4 channels, pointing to areas responsible for motor processing [11]. It is becoming increasingly clear that speech perception is at least partially located in the motor areas, especially under less-than-optimal listening conditions. This cruciality of the C4 channel was similarly seen in the Theta band and is in line with prior research evidencing the important role played by the lower frequency bands in general and Beta band coupling in particular [60]. Hence, inefficient phase locking in the auditory cortex may affect visual and motor processing development, which may in turn cause some of the visual, motor and attentional difficulties seen in DD [61].

It should be noted, however, that the C3-C4 interaction is mostly relevant for the sources and is not important for either the C3 for the sinks or, as a result, for the total activity. Meanwhile, the causal activity in the Beta band is different in the occipital area in the sinks scenario, and it is remarkably different in the frontal area, especially in FP1 for all three scenarios and in the F3 channel for the sinks scenario.

In the Gamma band the activity is higher than in the Theta band for maximum values, although the occipital area shows more concentrated activity among the causality relations, as Fig. 5 shows. Nevertheless, the effect is different between the control and dyslexic groups, whereby the participants with DD show higher activity for the sinks, which increases their total activity.

For the sources, the channels with the most explicit differences are FC1 and, more generally, TP9 in the left temporal area. In the case of sinks, this is also an important channel, although O1 and, as highlighted above, C3 also play a role.

Meanwhile, in the Gamma band, despite the discrepancies between the dorsal and ventral pathways, the latter offers the main difference for the classification of TP9 for both sources and sinks. FC1 is linked to sources and C3 to sinks, suggesting a significant cause-effect relationship, albeit with potentially less activity in the dyslexic group, facilitating classification.

We can confirm that the classifier performs better in the Theta and Gamma bands, which can evidence atypical oscillatory differences based on both speech and non-speech stimuli [56]. According to Leong's models [62], the slower rates (Delta and Theta) temporally constrain entrainment at the faster rates, such as Gamma.

Lehongre et al. [65] contended that the oscillatory nesting seen between the Theta/Delta phase and the Gamma power [63][64] offers a way to integrate information at the phonemic (Gamma) rate into the syllabic rate.

Meanwhile, the integration of the various acoustic features that contribute to the same phoneme being perceived may be hindered by impairments in the phase locking by Theta generators. Otherwise, flaws in certain Theta mechanisms could influence the development of the phonological system, which thus tends to code information bilaterally with the Gamma oscillations independently and then link them perceptually with the Theta oscillator output. In this case, the impaired phase locking of the right hemisphere Theta oscillatory networks causes difficulties with lower frequency modulations [17][66].

In addition, the spontaneous oscillatory neural activity identified in the auditory cortex in both the Theta and Gamma bands is known to associate with spontaneous activity in the visual and premotor areas [66].

A bilateral Gamma sampling network codes the signal's higher frequency modulations, thereby facilitating temporal integration at the phonetic (i.e., phoneme) scale. If we apply this model to DD, it is indicated that impaired processing at the syllable level (i.e., less efficient Theta phase locking) occurs alongside unimpaired Gamma sampling, meaning more weight is assigned to phonetic feature information during phonological development. Hence, as is the case in typical infant development, children with DD may have sensitivity to all phonetic contrasts of human languages [67].

Leong and Goswami [62] found that learners with DD show a preference for different phase alignment between amplitude modulations (AMs) when these respectively convey syllable and phoneme information (Theta and Gamma-AMs). A different phase locking *angle* suggests a discrepancy in the integration of speech information that arrives at a temporal rate different to that of the final perception of the speech [14]. Our results concerning the interaction between the Theta and Gamma bands support this.

Finally, our results also seem to confirm that the dyslexic brain is less efficient at encoding the amplitude modulation hierarchy's highest levels, i.e. those bearing information on the prosodic-syllabic structure, leading to cascade effects that impact the encoding of the phonological structure's levels nested within the Delta band, such as the syllable-level (Theta band) and phoneme-level (Gamma band) AM information.

Importantly, our results have been validated using a demanding permutation test, with the aim of ensuring that the results are not coincidental, despite the medium sample size.

## 6. Conclusion and future works

Our results support the main assumption of the TSF that DD involves a specific deficit in the low-frequency phase locking mechanisms in the auditory cortex, thereby potentially affecting phonological development [56].

In confirmation of this, we find an anomaly that emerges primarily in the causal relationships of channels that function as sinks, which is significantly more pronounced than when only the total activity is considered. Hence, it is reasonable to consider a division into Granger-causing or Granger-caused relationships. This, in turn, suggests that the main differences contributing to DD emerge when certain brain areas must function as receptors in the interactions between channels.

Furthermore, our results are in line with previous research, which has already detected an anomaly in the right-lateralized Theta band. We have clearly identified this here across all three scenarios (sources, sinks, total activity).

We also find confirmation for the higher brain activity in learners with DD, although differences are more significant for the sinks in the Theta and Gamma bands, in turn leading to more total activity. The highest classifier performance (accuracy and AUC) is hereby found in the sink scenario. For the Beta band, the difference in activity is more consistent across all three scenarios. The classifier also performs well for the Beta band in all three scenarios, with few differences observed, thereby confirming the important role played by this band in the sensorimotor coding of speech.

The results reflect the causal activity generated in the brain subjected to prosodic-syllabic stimulus at 4.8 Hz. Consequently, future work could consider the Granger causality relationships in the phases across channels and bands using higher frequency stimuli to stimulate syllabic-phonetic and phonetic activity.

## Acknowledgements

This work was supported by projects PGC2018-098813-B-C32 (Spanish "Ministerio de Ciencia, Innovación y Universidades"), UMA20-FEDERJA-086 (Consejería de econnomía y conocimiento,Junta de Andalucía) and by European Regional Development Funds (ERDF).